\documentstyle[pre,aps,tighten,epsf,floats]{revtex}
\begin{document}
\renewcommand
\baselinestretch{2}
\large
%\flushbottom
\title{\LARGE Mobility and stochastic resonance in spatially 
inhomogeneous systems \vspace{0.5in}}
\author{\large Debasis Dan,$^{1,*}$ Mangal C. Mahato,$^{2}$ and 
A. M. Jayannavar$^{1,\dagger }$ \vspace{0.2in}}
\address{\large $^1$ Institute of Physics, Sachivalaya Marg, 
Bhubaneswar 751005, India} 
\address{\large $^2$ Department of Physics, Guru Ghasidas University,
 Bilaspur 495009, India}
\maketitle
\begin{abstract}
\vspace{1.0in}
\large

		The mobility of an overdamped particle, in a periodic potential
tilted by a constant external field and moving in a medium with periodic 
friction coefficient
is examined. When the potential and the friction coefficient have the same 
periodicity
but have a phase difference, the mobility shows many interesting features as a
function of the applied force, the temperature, etc. The mobility shows 
stochastic resonance even for constant applied force, an issue of much recent
interest. The mobility also exhibits a resonance like phenomenon as a function
of the field strength and noise induced slowing down of the particle in an 
\vspace{0.2in}
appropriate parameter regime.\\
P. A. C. S: 82.20.Mj; 05.40.+j; 05.60.+w; 02.50.Ey 
\end{abstract}
%\begin{multicols}{2}
%\narrowtext
\newpage
\section{Introduction}

	The simple phenomenon of stochastic resonance ( SR) has been an active 
field of research for more than one and a half decades. Its appeal to natural
processes has generated a lot of interest among workers from various branches 
of science and engineering\cite{Gam}. This interest is largely because of the 
fact that 
in this phenomenon noise plays a useful role. The presence of noise helps in
improving the quality of response of a system subject to an alternating 
external field. The conventional SR 
is all about the( output signal-to-noise ratio) optimization of an input
signal by noisy nonlinear systems by suitably tuning the noise level. It 
is reflected as a peak in the output signal-to-noise ratio as a function 
of input
noise strength. To observe SR one requires three basic ingredients (i) an 
energetic activation barrier, (ii) a weak coherent input periodic signal, 
and (iii)
 a source of noise. Attempts are being made to 
reduce the number of constraints for the realization of SR. In our present
work, we are interested in the response optimization, in the presence of a
non-oscillating external signal.

		There was a suggestion that SR could be observed in the drift 
velocity of an overdamped Brownian particle in a tilted periodic potential 
\cite{Hu}.
The suggestion was an important one for it sought to do away with the 
oscillating 
input signal, considered to be an essential ingredient for the observability of 
the phenomenon of SR, and replace it with a constant uniform external field.
Though the suggestion has now been proved to be incorrect \cite{Git,Cas} 
for overdamped particles, 
it is argued to be correct for underdamped particles \cite{Mar} where 
the inertia of the 
particles could act as surrogate to the external oscillating field in 
a periodic potential.
The present work shows that SR can indeed be observed in the drift velocity 
of overdamped particles in a periodic potential subjected to non-oscillating
uniform constant field but in a medium where the particles experience spatially 
periodic frictional drag.

 The variation of space dependent friction $\eta(q)$ ( in the absence of 
spatial variations in temperature) influences the dynamics of the particles in a
potential field and helps the system to approach towards its equilibrium or
steady state. The relative stability of the competing states is governed by 
the individual Boltzmann factor, i.e inhomogeneity in friction does not affect
the stationary ( equilibrium) properties of a system but it does affect the
dynamical ( nonequilibrium) properties such as the relaxation rates. 
In contrast, 
temperature inhomogeneity, changes the relative stability of the otherwise 
locally stable states ( thereby creating new steady states). Moreover, the 
relative stability of the system in two different local minima depends 
sensitively on the temperature profile along the entire pathway connecting 
the two local minima. Temperature nonuniformity
 together with spatially varying friction coefficient or alone has been shown 
to give rise to a net current in a periodic potential
\cite{But,Kam,Mill,Jay,Par,Talk} in the absence of an applied bias.
 Spatially varying 
friction coefficient, $\eta(q)$ of the medium, alone, however cannot give the
net probability current $j$ in such a potential. But when, in addition, a 
constant force $F$ is applied a finite net current may result at a constant
temperature $T$ of the medium. We consider a sinusoidal potential with period
$2\pi$ and a sinusoidally varying model friction coefficient of the same period
 but with a phase difference $\phi$ and modulating amplitude $\lambda$. The 
mobility \cite{But,Ris}  $\mu$ ( $=\: <\frac{dq}{dt}>/F = \frac{2\pi j}{F}$,
 where $<\frac{dq}{dt}>$
is the average drift velocity of the particles) shows very interesting behaviour
in the parameter space of ($T,F,\lambda,\phi$). Apart from the above mentioned 
SR as a function of noise, the mobility also exhibits resonance like phenomenon
as a function of $F$ and the phenomenon of noise-induced stability etc.

		Before we present the results of our calculation in section 3 we
discuss the necessary formalism in section 2. We summarize our work 
in the last section 4, with discussion. 

\section{The mobility}

		The motion of an overdamped particle, in a potential $V( q)$
and subject to a space dependent friction coefficient $\eta( q)$ and an 
additional 
constant force field $F$, at temperature $T$, is described by the 
Langevin equation
\cite{Jay,Par,Ris,Mah}.
\begin{equation}
\frac{dq}{dt} = -\frac{(V^{\prime}(q)-F)}{\eta(q)} - k_{B}T\frac{\eta^{\prime}
(q)}
{\eta(q)^2}  + \sqrt{\frac{k_{B}T}{\eta(q)}}\epsilon (t)        .
\end{equation}
		Here $\epsilon(t)$ is the randomly fluctuating Gaussian noise
term with the property 
\begin{center}
$<\epsilon(t)> = 0$   ,\\
$<\epsilon(t) \epsilon(t_{1})> = 2\delta (t-t_{1})$ , \\
\end{center}
where $<..>$ denotes average over the distribution of the fluctuating quantity
$\epsilon (t)$. The primes in Eqn. (1) denote derivative with respect to the
space variable $q$. Eqn. (1) has been derived earlier using microscopic 
treatment of system-bath coupling \cite{Par,Mah}. It should be noted 
that the above equation 
involves a multiplicative noise with an additional temperature dependent drift
term. The additional term turns out to be essential in order for the 
system to approach the
correct thermal equilibrium state. The motion is equivalently described 
by the Fokker-Planck
equation \cite{Kam,Par,Mah,Kam2}. 
%%%%%%%%%
%%%%%%%%%%
\begin{equation}
\frac{\partial P(q,t)}{\partial t}  =  \frac{ \partial}{\partial q}  
\frac{1}{\eta (
q)}[k_{B}T \frac {\partial P(q,t)}{\partial q}  +  (V^{\prime}(q)-F)P(q,t)] .
\end{equation}
%%%%%%%%%%%%
%%%%%%%%%%%%
The stationary probability current $j$, for periodic $V( q)$ and $\eta(q)$ 
with the same
periodicity $2 \pi$, is given as \cite{But,Ris}
%%%%%%%%%
%%%%%%%%%%
\begin{equation}
 j  =  \frac{k_{B}T(1-e^{\frac{-2\pi F}{k_{B}T}})}{\int_{0}^{2\pi} 
e^{\frac{-V(y)+Fy}{k_{B}T}}dy \int_ {y}^{y
+2\pi}\eta (x) e^{\frac{V(x)-Fx}{k_{B}T}}dx}     ,
\end{equation}
%%%%%%%%%%%%%
%%%%%%%%%%
and the corresponding  mobility \cite{But,Ris} $\mu = \frac{2\pi j}{F}$. 
Clearly $j \rightarrow 0$ as $F \rightarrow 0$, but $\mu$ remains finite	
for finite temperature $T$. The mobility in this linear response regime 
\cite{Ris} is
given by 
\begin{equation}
\lim_{F \rightarrow 0} \mu = \frac{(2 \pi)^{2}}{\int^{2\pi}_{0} dy 
e^{\frac{-V(y)}{k_{B}T}}\int^{y+2\pi}_{y}dx\eta(x) e^{\frac{V(x)}{k_{B}T}}} .
\end{equation}
 Notice that the above expression for $\mu$ involves a combination of $\eta(x)$
 and $V(x)$ which contribute in different ways and cannot be accounted for 
by a single
"effective" potential. We would like to point out that a particle in a medium,
with space dependent temperature profile, experiences an effective potential
( or net force). This in turn, is able to shift the stable points and introduces
noise induced transitions\cite{But}. In our present work we emphasize 
that the space dependent
friction influences the kinetic properties of the system and unlike in the case
of space-dependent temperature we do not have a net additional force 
in the steady state.

                      In the absence of any external periodic potential the
drift velocity is linear in $F$ and the corresponding mobility is given by 
$\mu = \frac{1}{\bar{\eta}}$, where
$\bar{\eta} = \frac{1}{2 \pi}\int^{y+2\pi}_{y} \eta(x) dx$, is the 
average value of the friction
coefficient over a period. The above result follows from straight forward
mathematical analysis. From Eqn. (3) one can see that the magnitude of $\mu$
depends sensitively on the potential and the frictional profile over the 
entire period. And it can be much larger or smaller than the mobility 
of a particle
moving in a homogeneous medium with friction coefficient equal to $\bar{\eta}$.
Depending on the temperature and the applied field, mobility can in fact be much 
larger than the asymptotic mobility \\( with respect to $F$ and $T$). 
This implies  that at some
particular values of $F$ and $T$ the mobility, in the absence of periodic 
potential is much smaller than in the presence of periodic potential. This 
counter-intuitive result emerges because of the complex manner in which 
the phase shift
between the potential and the frictional profile, the amplitude of the friction
coefficient, etc, influence the mobility.
		We consider for our calculation,
\begin{eqnarray}
V(q) & = & -\sin q  ,\\
\eta(q) & = & \eta_{0}(1-\lambda \sin (q+\phi)) ,
\end{eqnarray} 
with $0 \leq \lambda < 1$ so that $\bar{\eta} = \eta _{0}$. One can easily see 
that for $F < 1$, as $T\rightarrow 0$, $\mu \rightarrow 0$.
Also, as $F,\: T \rightarrow \infty,\: \mu \rightarrow \frac{1}{\eta _{0}}$. 
Moreover,
 $j(F) \neq -j(-F)$, except when $\phi = 0$ and $\pi$, and hence
mobility is asymmetric with respect to the force. For intermediate values of $T$
and $F$ one needs to evaluate the double integral in the denominator of Eqn. (3)
numerically \cite{Alan}and explore the variation of $\mu$ :
\begin{equation}
\eta_{0}\mu = \frac{2\pi k_{B}T(1-e^{\frac{-2\pi F}{k_{B}T}})}{F\int^{2\pi}_{0}
 e^{\frac{(\sin y+Fy)}{k_{B}T}}dy \int^{y+2\pi}_{y} (1-\lambda \sin (x+\phi))
 e^{\frac{-\sin x-Fx}{k_{B}T}}dx}  \;.
\end{equation}
		We calculate $\mu$ in various sections of the parameter space
of $(T, F, \lambda, \phi)$. Henceforth we have scaled mobility, temperature and
force to dimensionless units as in Ref. \cite{Ris}.

\section{RESULTS.}

	The mobility $\mu$ shows many interesting features even at $T=0$.
 At $T = 0$, following Ref.\cite{Ris} closely, $\mu$ remains zero 
for $|F|\leq 1$ for all
values of $\phi$ and $\lambda$. However, $\phi$ and $\lambda$ play 
important role
for $|F| > 1$ and the mobility is given by
\begin{equation}
\eta_{0}\mu = \frac{\frac{1}{F}(1-\frac{1}{F^{2}})^{1/2}}{\frac{1}{F}+
\lambda(1-\sqrt{1-\frac{1}{F^{2}}})\sin \phi} .
\end{equation}
		For applied fields smaller than the critical field ($F=1$) the
particle remains in a locked state or confined to the local minima of the 
potential. For $F > 1$ (i.e, when the potential barrier for motion does not 
exist)  the particle moves down the potential slope or will be
found in a running state \cite{Mar,Ris}. Observe that at $T=0$, 
$\mu(F,\phi) = \mu (-F,-\phi)$ and this
happens to be true for $T\neq 0$ as well. We may roughly divide the full range
of $\phi$ (0-2$\pi$) into two regions where $\mu$ shows qualitatively distinct
features [Fig. (1)].

From Fig. (1a) we observe that for $F > 1$, $0 \leq \phi \leq \pi$, 
$\eta_{0}\mu$ 
always remain less than one but approaches one asymptotically as 
$F \rightarrow \infty$.
Fig. (1b) shows that in the range $\pi < \phi < 2\pi$,  $\eta_{0}\mu$ could even
be larger than one for large $\lambda$ and exhibits a maximum. 
This occurrence of maxima is a new 
feature and could only be ascribed to the space dependence of 
friction coefficient
$\eta(q)$. 
It is to be noted that the value of the mobility
at the maxima is always larger than the asymptotic mobility 
($\frac {1}{\bar{\eta}}$). In Fig. (2) we have given a surface plot 
of the mobility obtained from 
Eqn. (8) as a function of $F$ and $\phi$. This figure clearly 
shows that there is a small region in  $\phi$, around
$1.2 \pi$ where the maxima can be observed, elsewhere mobility monotonically 
increases with $F$. In contrast, in a medium with a space independent 
friction , the mobility in the high friction limit monotonically increases
with $F$ for all values of $T$ and asymptotically goes to one \cite{Ris}.
For given $F > 1$, however from Eqn. (8), maximum occurs at $\phi = \frac{3\pi}
{2}$ and minimum at $\phi = \frac{\pi}{2}$.
 It should be noted, that the space dependence of the friction 
coefficient does not alter the threshold value ($F = 1$) for nonzero mobility.

As the temperature $T$ is increased from zero, the thermal
fluctuations make the current $j$ nonzero even for $|F| < 1$, when the barrier
to free passage motion is nonzero. For $T \neq 0$, as $F \rightarrow 0, j
\rightarrow 0$ linearly with $F$ but the mobility remains finite [Eqn. (4)].
For a given temperature $T \neq 0$
and when the field is very high, the effect of the periodic potential vanishes
and consequently $\eta_{0}\mu \rightarrow 1$. However in 
the small $F$ regime $\eta_{0}\mu$ can be much larger than one, leading to a 
peak as a function of $F$ depending on the parameter values of $\phi , \lambda 
,  T$. 
The resonance feature as a function of $F$ can also be observed
at finite temperatures, as shown in Fig. (3).

		Let us now examine the variation of $\eta_{0}\mu$ as a function 
of the modulation parameter $\lambda$ of the friction coefficient. 
Fig. (4) shows
the variation of $\eta_{0}\mu$ for $\phi =0 $ at fixed $T = 2.0$ and Fig. (5) 
for $\phi = \pi$ at the same $T$ ( For $\phi =0$ and $\pi,\;\eta_{0}\mu$ is
symmetric about $F = 0$).  When $ V(q)$ and $\eta (q) $ are in phase, 
the  mobility 
monotonically decreases with $\lambda$ ($F = 0$ having the least value 
at $\lambda = 1$
) whereas when they are in opposite phase mobility increases with 
$\lambda$. For $\phi=\pi$ and large $\lambda$ ( Fig. [5]), $\eta_{0}\mu$ becomes
larger than one for all values of $F$. 
Figures (4 and 5) essentially corroborate the observations 
made in Ref.\cite{Fal} that when $\phi = 0$ the friction coefficient 
is the largest 
at the positions where the particle has the largest acquired velocity due to the
potential, and it is smallest where the particle has the smallest velocity.
 Thus the damping term has the severest effect of slowing down the particle.
When $\phi = \pi$ the situation is just the opposite. In this case 
the frictional 
effect is the least. In fact it has a positive effect and as can be seen from
Fig. (5) the mobility $\eta_{0}\mu$ even becomes larger than one. For any other 
$\phi$, however, the situation is complicated to analyze as simply. To emphasize
this we have given a surface plot of $\eta_{0}\mu$ with $\lambda$ and $\phi$
in Fig. (6) for $T = 2.0$ and $F = 0.8$. Mobility increases monotonically with
$\lambda$ in a region for $\phi$ around $1.1 \pi$, and elsewhere it 
decreases monotonically.

		We now discuss the variation of $\eta_{0}\mu$ as a function of 
temperature $T$ ( or the noise strength). At this point it is pertinent to note 
that there has been discussion in literature about the motion of an underdamped
particle, in tilted ( due to a constant force $F$) but otherwise 
periodic potentials,
being in two states depending on the value of $F$ and the damping constant
$\eta_{0}$ of the medium \cite{Mar,Ris}. The particle could be in a 
locked state ( remaining
confined to a local minimum of the potential) or in a running 
state ( corresponding
to a motion down the potential slope). The two states could also coexist even 
for $|F| < 1$ ( for periodic potential $V(q)$ as given in Eqn. (5) ). 
However, in
the overdamped case at $T = 0$ the particle can be either in the locked state 
(for $|F| < 1$) or in the running state( for $|F| > 1$). But as $T$ is increased
the particle incoherently hops from one stable \\ ( locked state) to another one
even for $|F| < 1$, leading to a net current. When in addition, we consider the
friction coefficient to be space dependent, the mobility as a 
function of temperature
shows many more interesting features depending on the values of $F, \lambda$ and
$\phi$. We confine our attention to large values of $\lambda$ ( but $< 1$), for
example, $\lambda = 0.9$.

              In the entire range of $\phi [0 < \phi <2\pi]$ we present results
for a few typical values of $\phi$. \\Fig. (7a) shows the nature of SR at $\phi
= 0.9 \pi$ for arbitrarily selected values of $F$. We see that for some 
values of
$F$, especially for $F < 1$, $\eta_{0} \mu$ becomes larger than one at 
intermediate values  of $T$  \\( the peaks appear at
larger $T$ for larger $F$) and asymptotically tends to one as $T \rightarrow
\infty$, for all values of $F$. We find SR even for $F \rightarrow 0$ for
$F \rightarrow 0$ and in fact the peak mobility is larger for $F < 1$ than for 
$F > 1$. Similar behaviour could be seen for $ \phi<0.9 \pi$ also. These peaks
for small $\phi$, however, are very broad. Fig. (7b) is for $\phi = 1.44 \pi$. 
In this figure the peaks are sharper for larger
values of $F$ ($F > 1$) and the peaks  disappear for $F < 1$.
The temperature $T (= T_{P})$ corresponding to the peak value of mobility is
much lower than that in the previous case ($\phi = 0.9 \pi$) and unlike 
the previous case it decreases and tends to zero as $F$ is increased.
Figs. (7a) and (7b) clearly exhibits the occurrence
of SR for $F < 1$ as well as for $F > 1$.  
It should be noted that we observe stochastic resonance for smaller values of 
$\lambda( <0.9)$ too. But  as $\lambda$ is decreased the peaks become less 
prominent and the range of $\phi$ over which the phenomena could be 
observed shrinks and finally disappears.

We now turn to examine the location of maxima $T_{P}$, as a function of system
parameters. In Fig. (8) we have plotted $T_{P}$ against $\phi$ for various 
values of $F$, the inset shows the variation of $T_{P}$ versus $F$ for some
typical values of $\phi$ as indicated on the figure. For small fields $F < 1$,
$T_{P}$ versus $\phi$ is a non-monotonic function exhibiting a minima, which
shift towards  higher values of $\phi$ as $F$ as increased. As we go away from
the minima on both sides, the resonance behaviour  becomes more broader. For
$\phi$ beyond $1.5 \pi$, mobility does not show a maxima( or $T_{P}$). For 
values of field larger than about $1.2$, $T_{P}$ decreases monotonically with
$\phi$. $T_{P}$ shows a monotonic or a non-monotonic behaviour as a 
function of $F$
as can be seen from the inset of Fig. (8). This behaviour is very sensitive to
material parameters.
In Fig. (9) we have given a surface plot of the mobility as a function 
of $\phi$ and $T$ for
 $F = 0.5$ and $\lambda = 0.9$. For values ranging from $\phi = 0.9 \pi$ to 
$1.44 \pi$ we obtain SR in the mobility as a function of temperature. It is
to be noted that this surface plot as well as all the earlier surface plots 
exhibit a maxima in the mobility as a function of $\phi$ for a given temperature
 $T$ and field strength $F$, on the other hand the mobility shows a 
monotonic behaviour
with $\lambda$ (see for example Fig. (6)). This seems to suggest that $\phi$ may
play the characteristic role of frequency in our model in the 
absence of additional external frequency signal \cite{Mar,Kim}.

Fig. (10a) is for $\phi = 1.5\pi$. For forces less than a
certain $F$ (closer to the critical value of $F = 1$) all curves begin 
with $\eta_{0}\mu$ less than one and gradually
increase with $T$ and asymptotically reach the value one. And for
larger forces the mobility always remain larger than $\eta_{0} \mu = 1$
and asymptotically reach one as $T \rightarrow \infty$. 
For large values of $F( > 1$), for
which the potential barrier for motion has already disappeared, the mobility 
decreases as $T$ is increased from $T = 0$. This implies that the motion of the
overdamped  particle becomes more sluggish than at $T = 0$, where one would
have expected the particle to become more mobile as $T$ is increased from zero.
Our observation indicates that the presence of noise actually slows down the
motion of deterministically overall unstable states in the appropriate range of
$F$ and $\phi$. This is somewhat akin to the phenomenon of 
noise-induced stability
of unstable states \cite{Day,Mat,Mah2,Mah3}. For the values of 
$F$ larger than the critical field,
the barrier to the motion of the particle disappears and in this case
( depending on the system parameters)
the noise slows down the motion between unstable states whereas for $F < 1$ the
noise helps to overcome the potential barrier for the passage of the particle.
And finally in Fig. (10b), for $\phi = 1.6 \pi$ , we observe 
some very interesting features that are
somewhat contrary to common expectations.
 Values of $F$ very close to ( but larger than the critical value one), 
the mobility attains a minimum
value and then begins to increase with $T$ and finally reaches 
$\eta_{0}\mu = 1$, 
asymptotically, as $T \rightarrow \infty$. In this case we observe a cross-over
 from the regime of noise-induced slowing down to the noise-enhanced mobility 
for a given $F$ as a function of the temperature (noise strength). 
All these interesting features ( some
 of them even counterintuitive) described above, results from a 
subtle combined effect
of the periodic space dependent friction and the periodic potential in the 
presence of a constant applied force. The phase difference $\phi$ between the
potential and the friction coefficient plays a crucial role in determining
their nature. 

\section{Summary and Discussion}

 We have theoretically studied the motion of an overdamped  particle in a
tilted( by a constant force) but otherwise sinusoidal potential 
in an inhomogeneous
medium. We considered the particle to experience a sinusoidally varying friction
 coefficient while in motion, but the temperature of the medium is kept uniform
. The behaviour of the motion of the particle characterized by its mobility,
is influenced by the phase difference between the potential function and the
friction coefficient in important ways. If the friction coefficient is small
where the potential is high, then the bias $F$ will influence the mobility 
differently than when the friction coefficient is large where potential is high.
By choosing the phase difference suitably
the mobility could be enhanced by tuning  the noise level \\( temperature)
of the system or lowered depending on the value of the phase difference. The 
former indicating the occurrence of stochastic resonance in the motion of the 
overdamped particle in a periodic potential with the application of a 
non-oscillating constant field.
The decrease of mobility, on the other hand, indicates that the motion 
could be made sluggish by enhancing the noise level \\( temperature). 
We have also observed 
the peaking behaviour in the mobility as a function of applied field. 
Of course, both 
these effects cannot be seen in a homogeneous medium with constant friction
coefficient and uniform temperature.

	At this point it is worthwhile to note that in the conventional
treatment on SR it has been established that SR occurs when Kramer's time 
of barrier passage roughly matches the ( dominant characteristics) time 
scale of the input ( externally applied) signal. But in the present work there
is no time varying externally applied input signal and hence the criterion
for the occurrence of SR is not satisfied.
However some recent works have shown that the SR can also occur in the
absence of an external periodic force, which is a consequence of the intrinsic
dynamics of the nonlinear system \cite{Cab,Hakn}. Such a behaviour has 
been described as 
autonomous SR. It may also be possible that the SR phenomena 
that we have discussed, 
may be a little closer to dither( or threshold SR) .

		In this system, that we study, the current is in general 
asymmetric with respect to the reversal of the direction of the applied force. 
This result has earlier been used to obtain a net unidirectional motion in the
presence of time oscillating field, but with net average force zero( i.e without
 an obvious bias)\cite{Zhao}.

\section*{ Acknowledgement \hfill}

MCM thanks the Institute of Physics, Bhubaneswar, for financial assistance
and
hospitality. MCM and AMJ acknowledge partial financial support from the
B. R. N. S. /  D. A. E. , India.\vspace{0.5in}

\hspace{1.8in} \Large REFERENCES \hfill

%\end{multicols}{2}

\newpage
\hspace{1.5in} \large  FIGURE CAPTIONS .\vspace{0.5in} 
\large

FIG. 1. Mobility $\eta_{0} \mu$ as a function of $F$ at $T=0$ for various 
values of $\lambda$ (all in dimensionless units ) for (a) 
$\phi = 0.9 \pi $,  (b) $\phi = 1.2 \pi $.

\vspace{.3cm}
FIG. 2. Surface plot of Mobility at $T = 0$ as a function of $F$ and $\phi$ at
 $\lambda = 0.9$. $\phi$ is in units of $2 \pi$.

\vspace{.3cm}
Fig. 3. Surface plot of mobility $\eta_{0}\mu$ as a function of applied bias $F$
and $\phi$ at $T = 0.8$ and $\lambda = 0.9$.  $\phi$ is in units of $2 \pi$.

\vspace{.3cm}
FIG. 4. Mobility $\eta_{0} \mu$ as a function of $\lambda$ at $T=2.0$ 
and $\phi = 0$ for various values
 of $F$ .

\vspace{.3cm}
FIG. 5. Mobility $\eta_{0} \mu$ as a function of $\lambda$ at $T=2.0$ 
and $\phi =
 \pi$ for various values
 of $F$ .

\vspace{.3cm}
FIG. 6. Surface mobility as a function of $\lambda$ and $\phi$ at $F = 0.5$ and
 $T = 2.0$, to highlight the monotonic behaviour of mobility with $\lambda$.

\vspace{.3cm}
FIG. 7. Mobility $\eta_{0} \mu$ as a function of temperature $T$ 
at $\lambda = 0.
9$ for various values
 of $F$ for (a) $\phi = 0.9 \pi$ (b) $\phi = 1.44 \pi$ . The insets highlight 
the maxima in the curves.

Fig. 8. Temperature corresponding to the peak mobility ($T_{P}$) as a function 
of $\phi$ ( in units of $2\pi$ ) for various values of $F$. The inset shows
the variation of $T_{P}$ as a function of  $F$ for various values of $\phi$.

Fig. 9.  Surface plot of mobility $\eta_{0}\mu$ as a function of temperature $T$
and $\phi$ at $F = 0.5$ and $\lambda = 0.9$. $\phi$ is in units of $2 \pi$.

FIG. 10. Mobility $\eta_{0} \mu$ as a function of temperature 
$T$ at $\lambda = 0.
9$ for various values
 of $F$ for (a) $\phi = 1.5 \pi $ (b)$ \phi = 1.6 \pi$. The inset of figure (b)
highlights the minima in mobility.

\vspace{.3cm}
%% 1 %%
\begin{figure}
  \centerline{\epsfysize=19cm \epsfbox{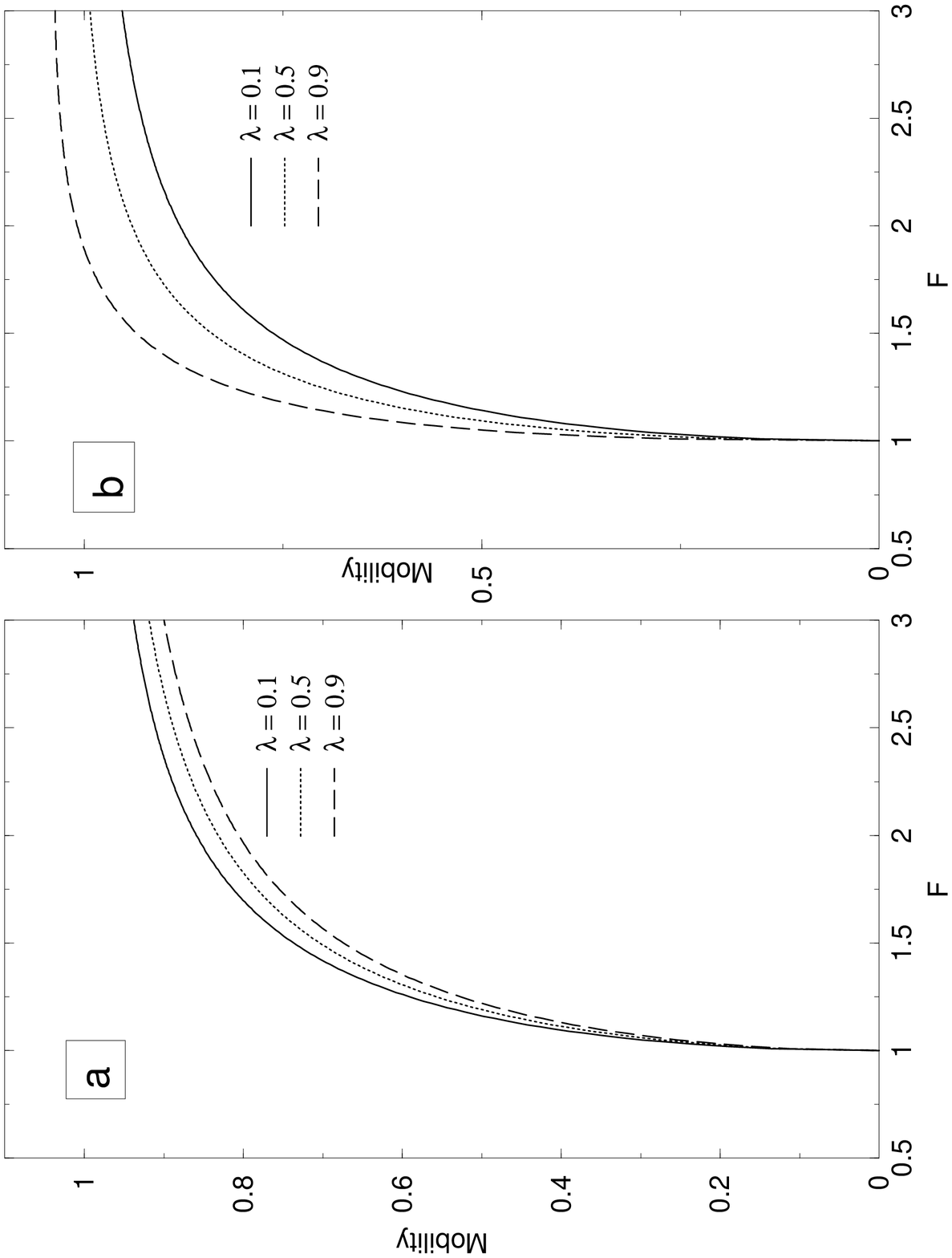}}
  \vspace{.5in}
  \caption{}
\end{figure}

%% 2 %%
\begin{figure}
  \centerline{\epsfysize=12cm \epsfbox{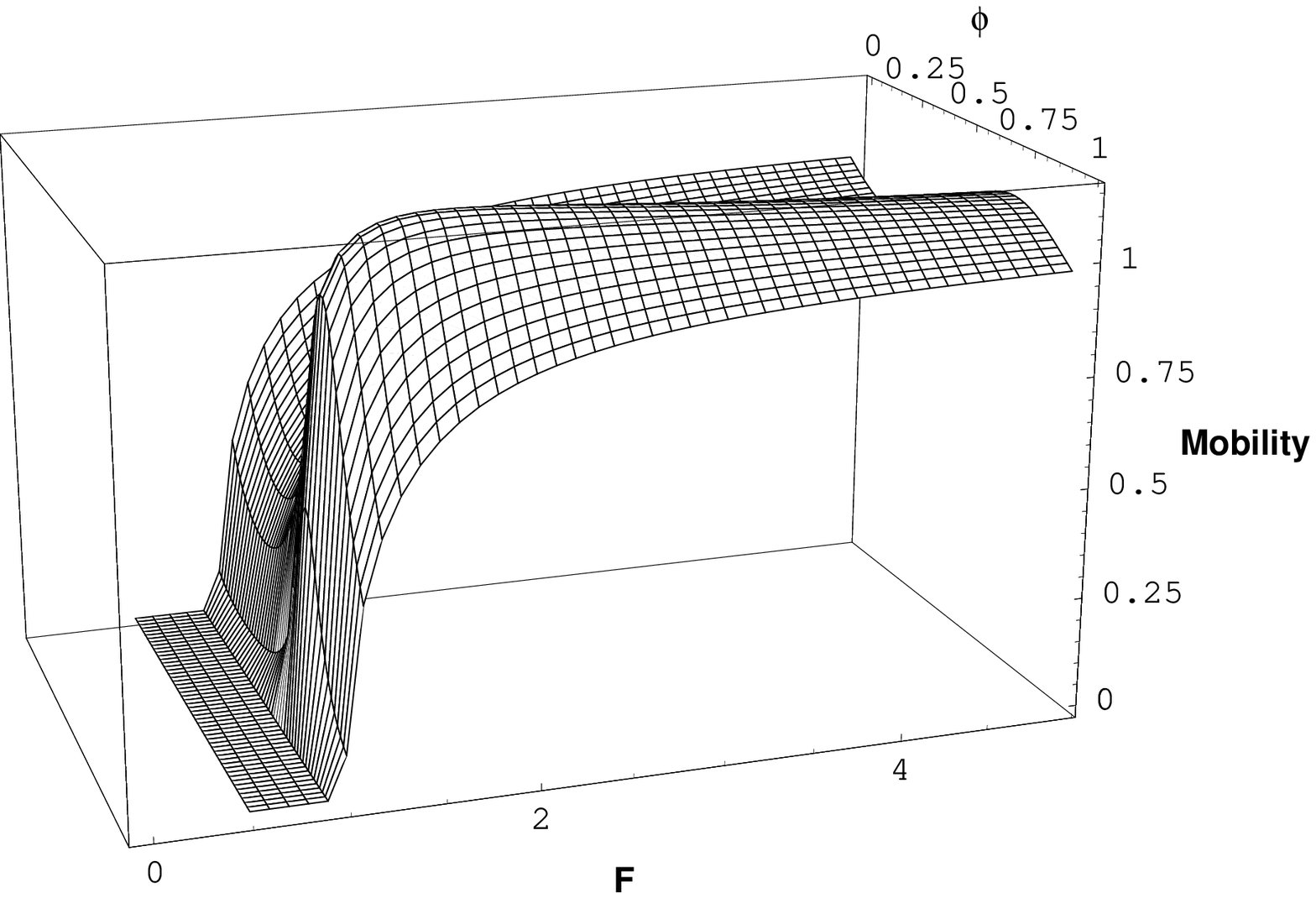}}
  \caption{}
\end{figure}

%% 3 %%
\begin{figure}
  \centerline{\epsfysize=11cm \epsfbox{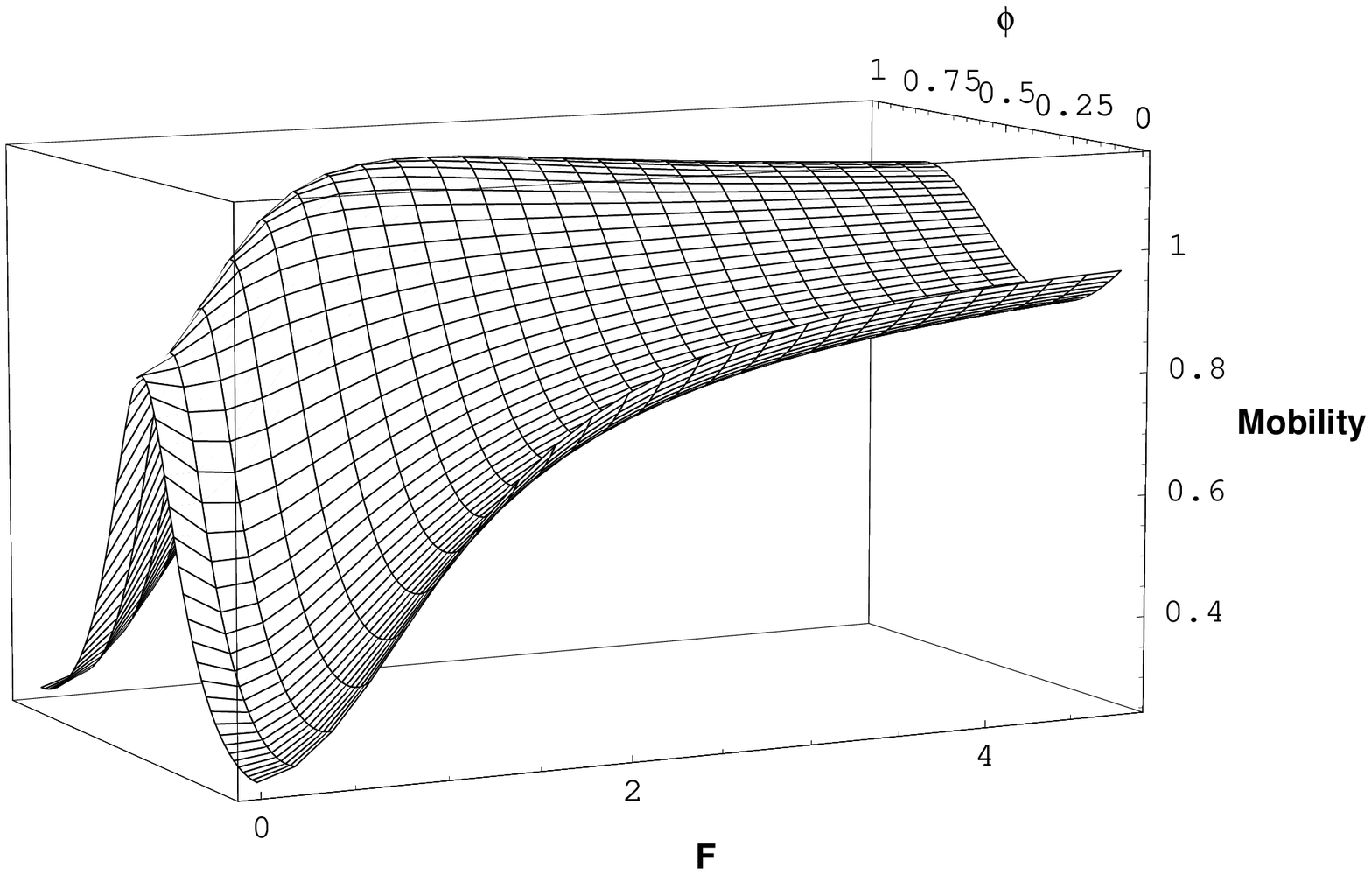}}
  \caption{}
\end{figure}

%% 4 %%
\begin{figure}
  \centerline{\epsfysize=19cm \epsfbox{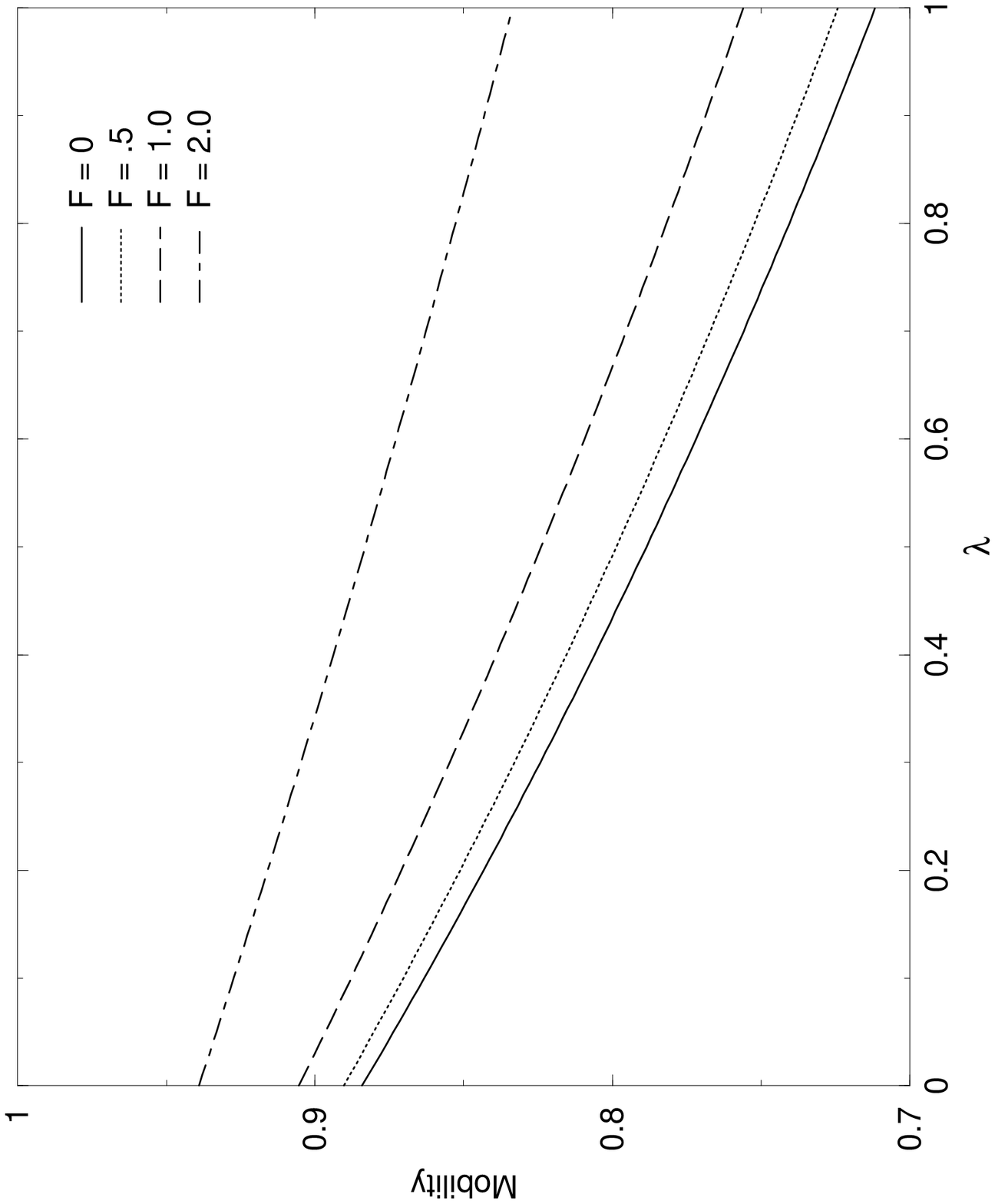}}
  \caption{}
\end{figure}

%% 5 %%
\begin{figure}
  \centerline{\epsfysize=19cm \epsfbox{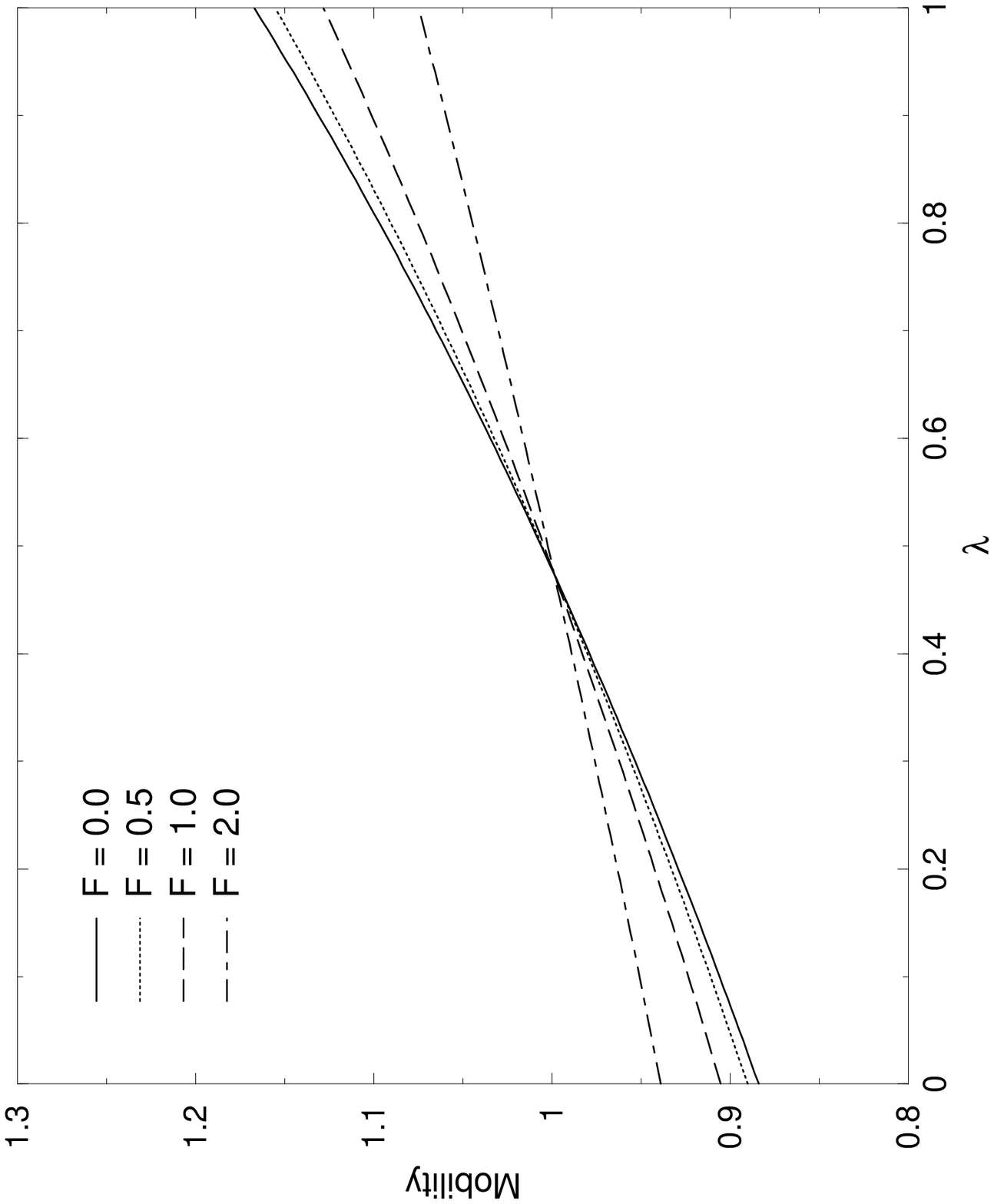}}
  \caption{}
\end{figure}

%% 6 %%
\begin{figure}
  \centerline{\epsfysize=11cm \epsfbox{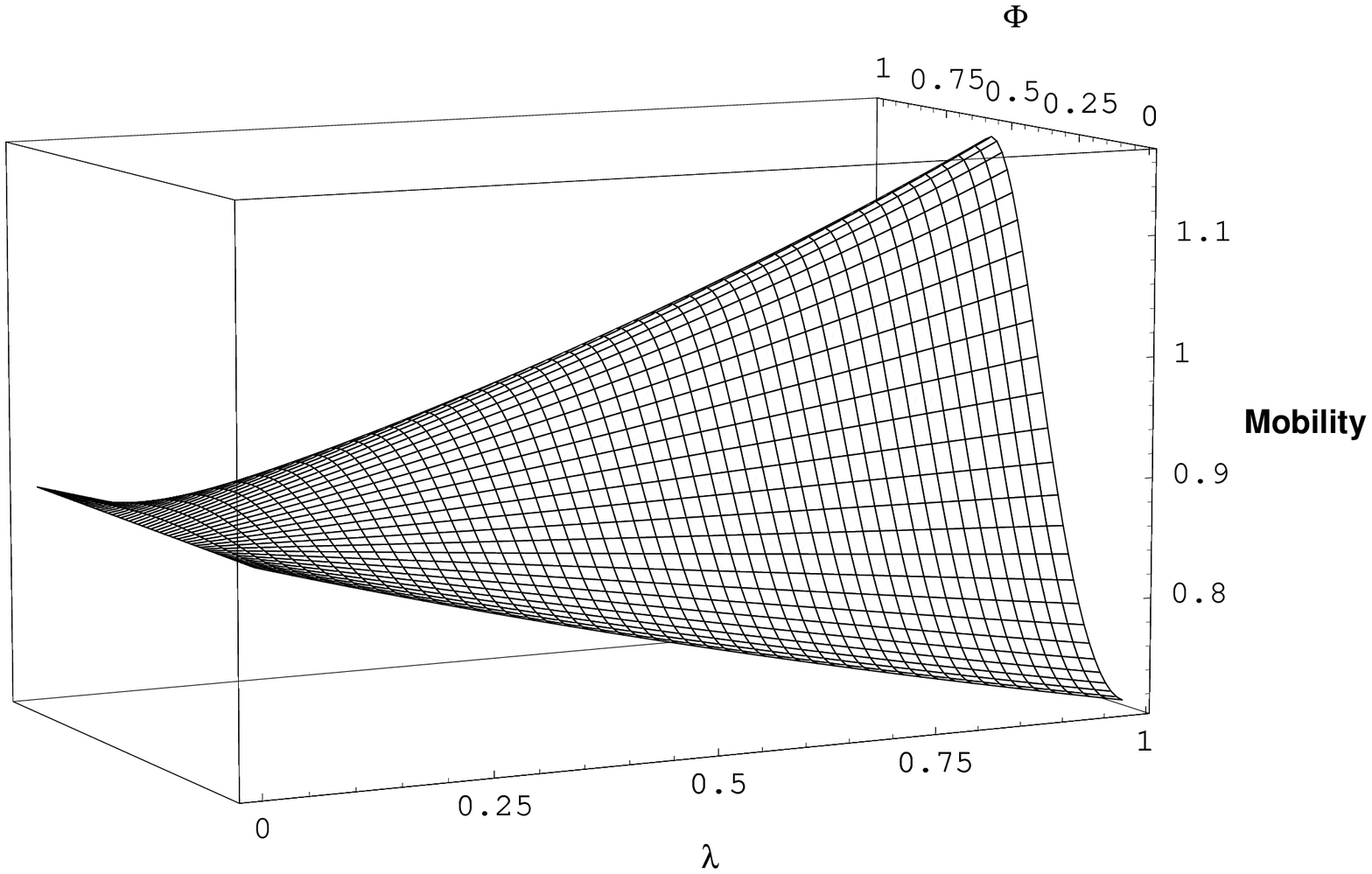}}
  \caption{}
\end{figure}

%% 7a %%
\begin{figure}
  \centerline{\epsfysize=19cm \epsfbox{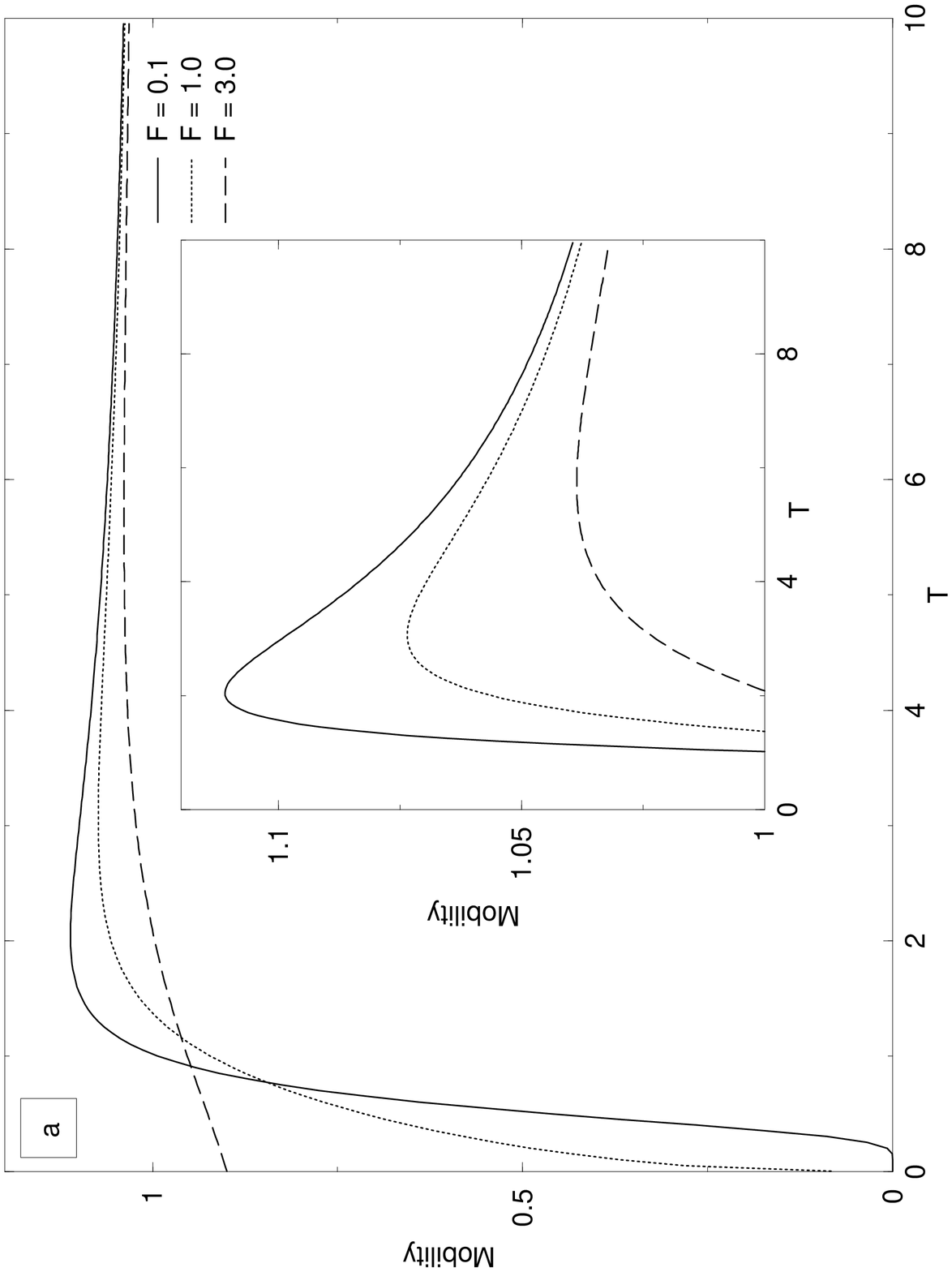}}
  \vspace{.5in}
  \caption{}
\end{figure}

%% 7b %%
\begin{figure}
  \setcounter{figure}{6}
  \centerline{\epsfysize=19cm \epsfbox{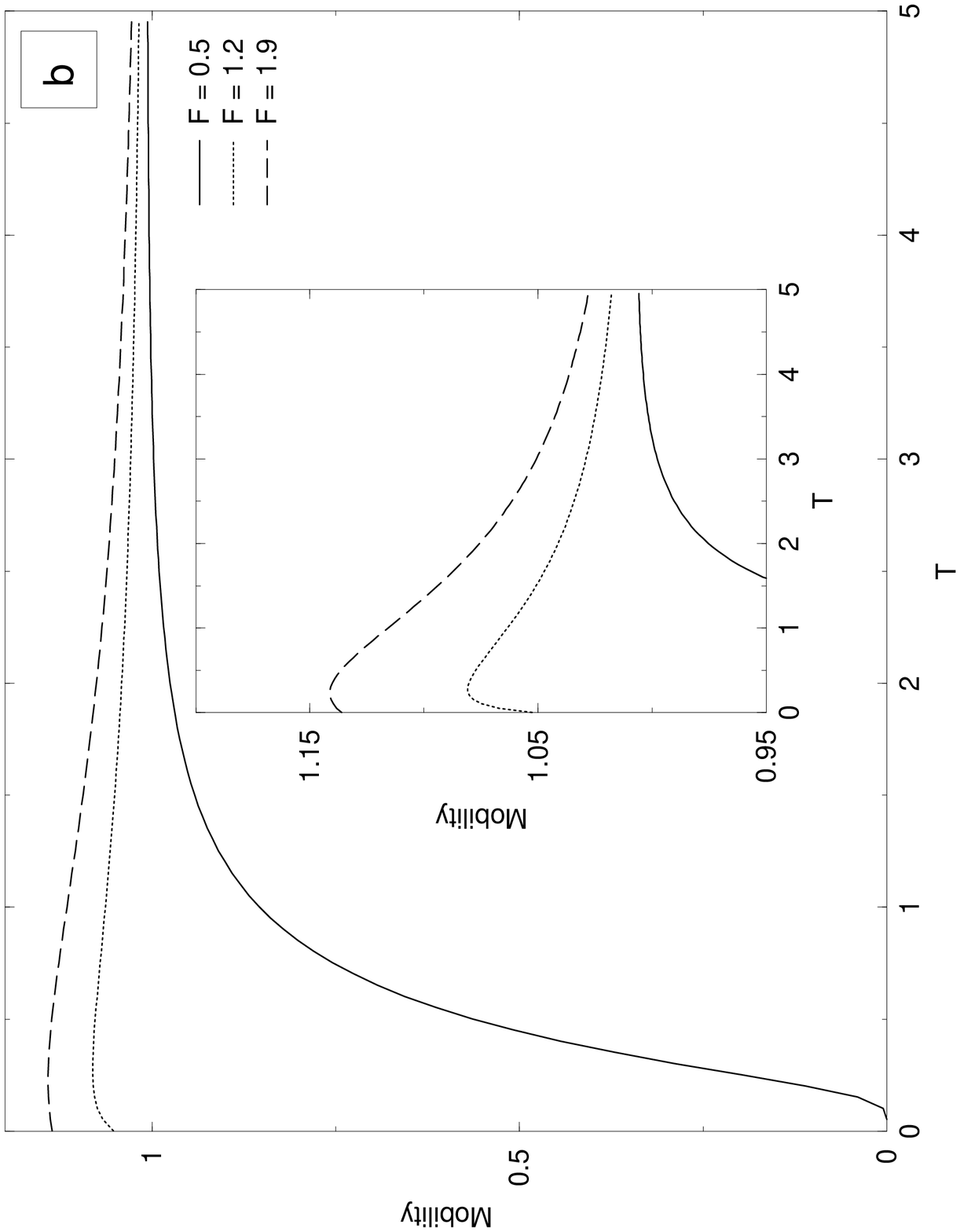}}
  \vspace{.5in}
  \caption{}
\end{figure}

%% 8 %%
\begin{figure}
\centerline{\epsfysize=20cm \epsfbox{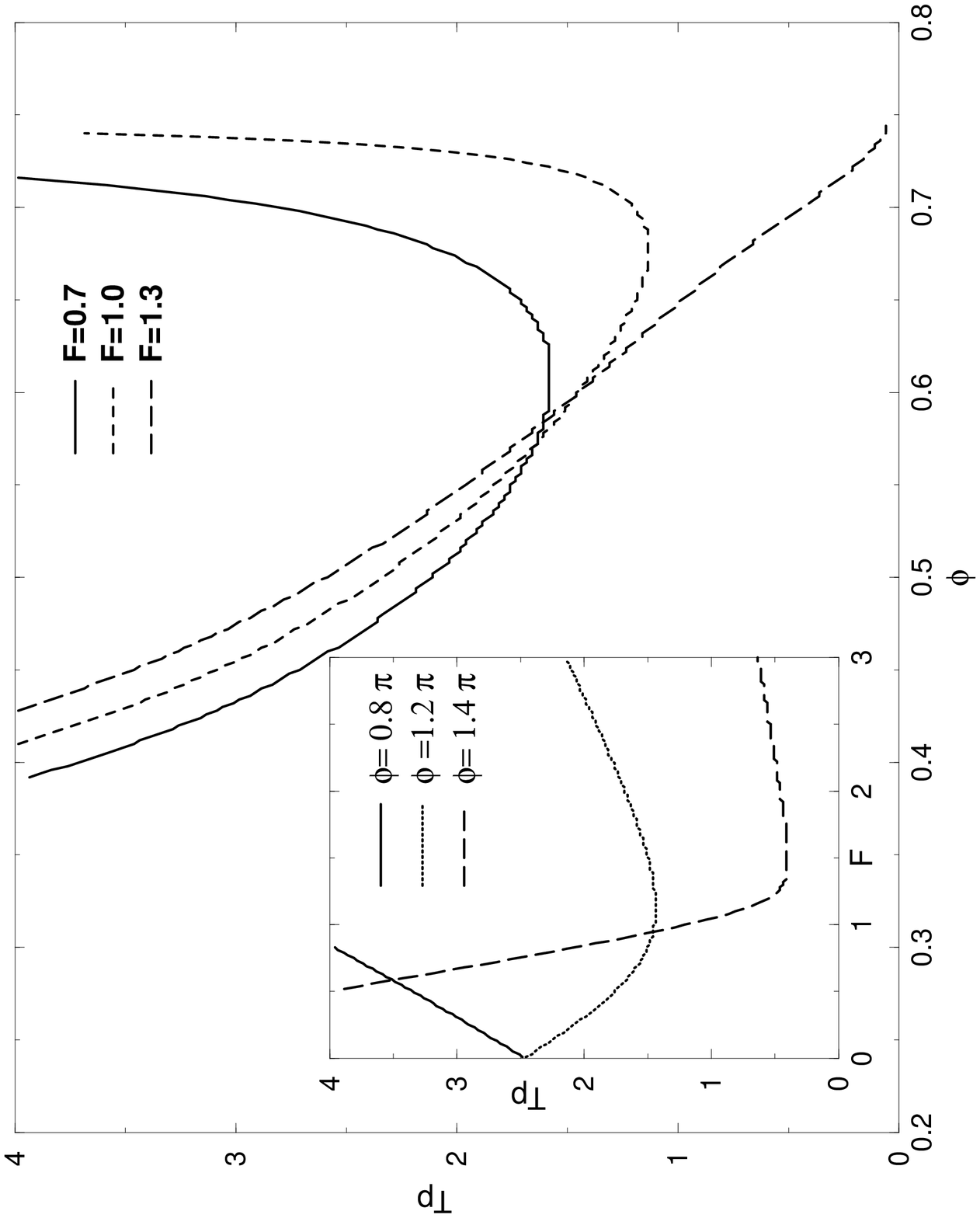}}
\vspace{.5in}
\caption{}
\end{figure}

%% 9 %%
\begin{figure}
  \centerline{\epsfysize=17cm \epsfbox{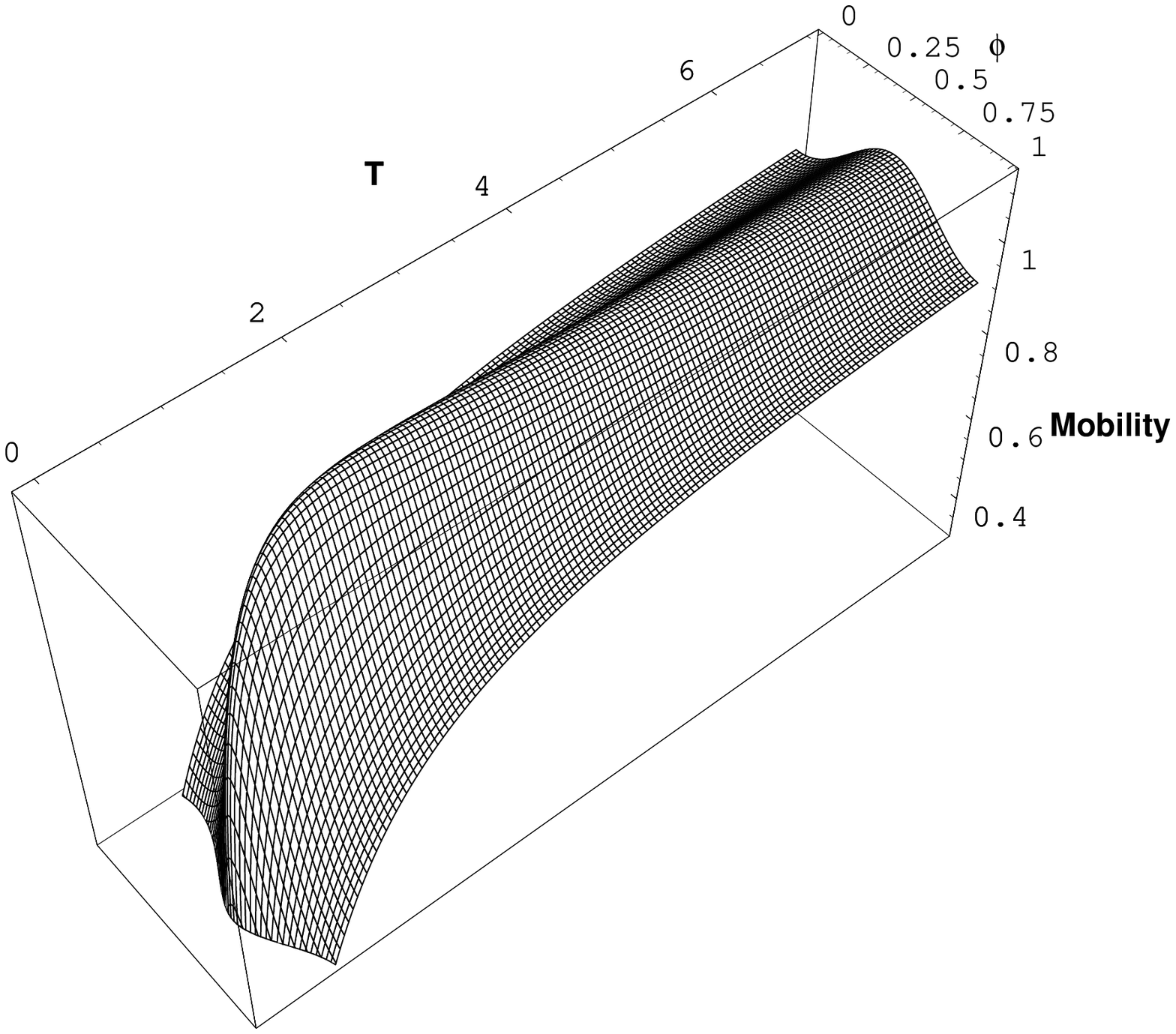}}
  \caption{}
\end{figure}

%% 10a %%
\begin{figure}
  \centerline{\epsfysize=19cm \epsfbox{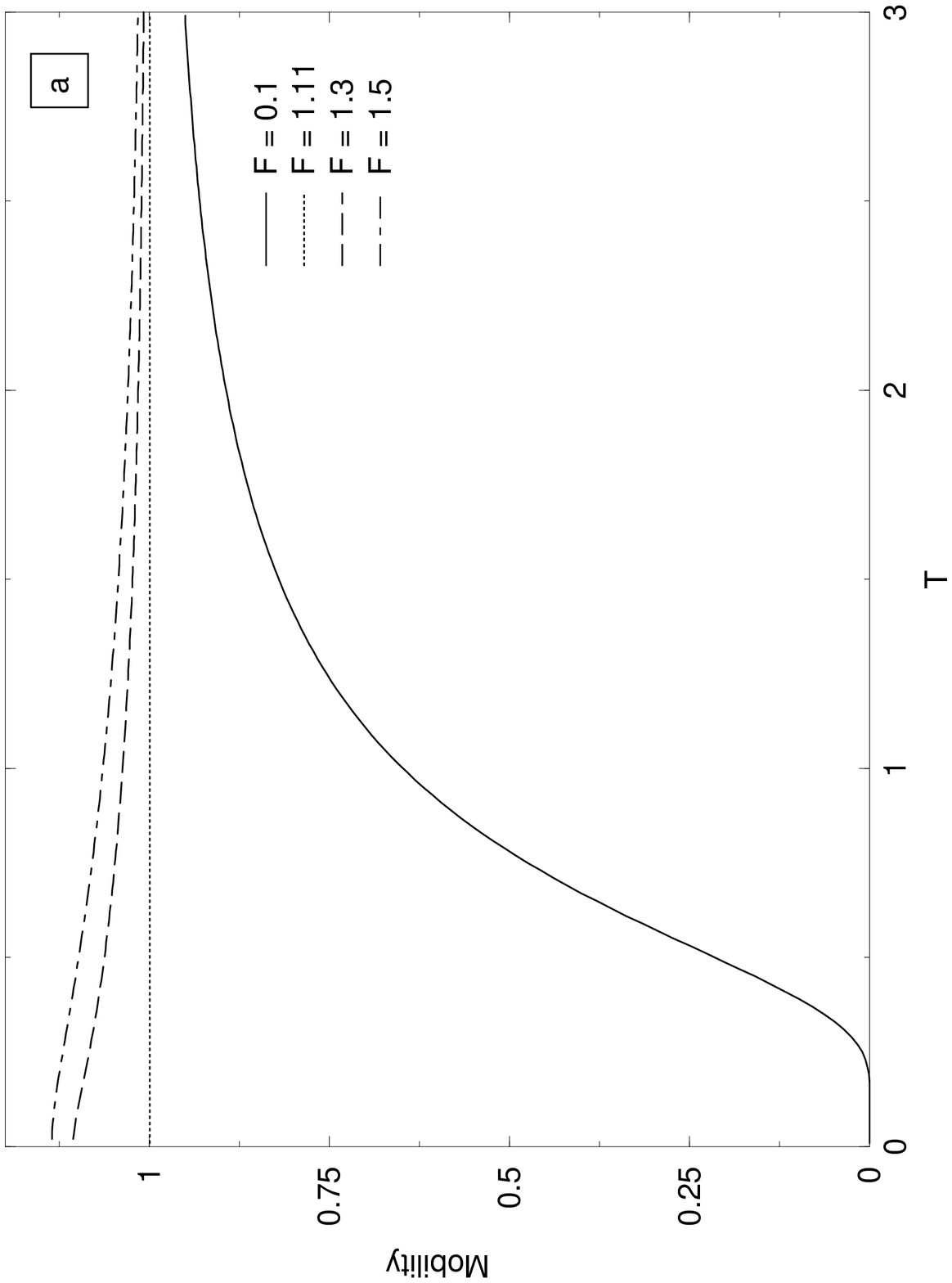}}
  \caption{}
\end{figure}

%% 10b %%
\begin{figure}
  \setcounter{figure}{8}
  \centerline{\epsfysize=19cm \epsfbox{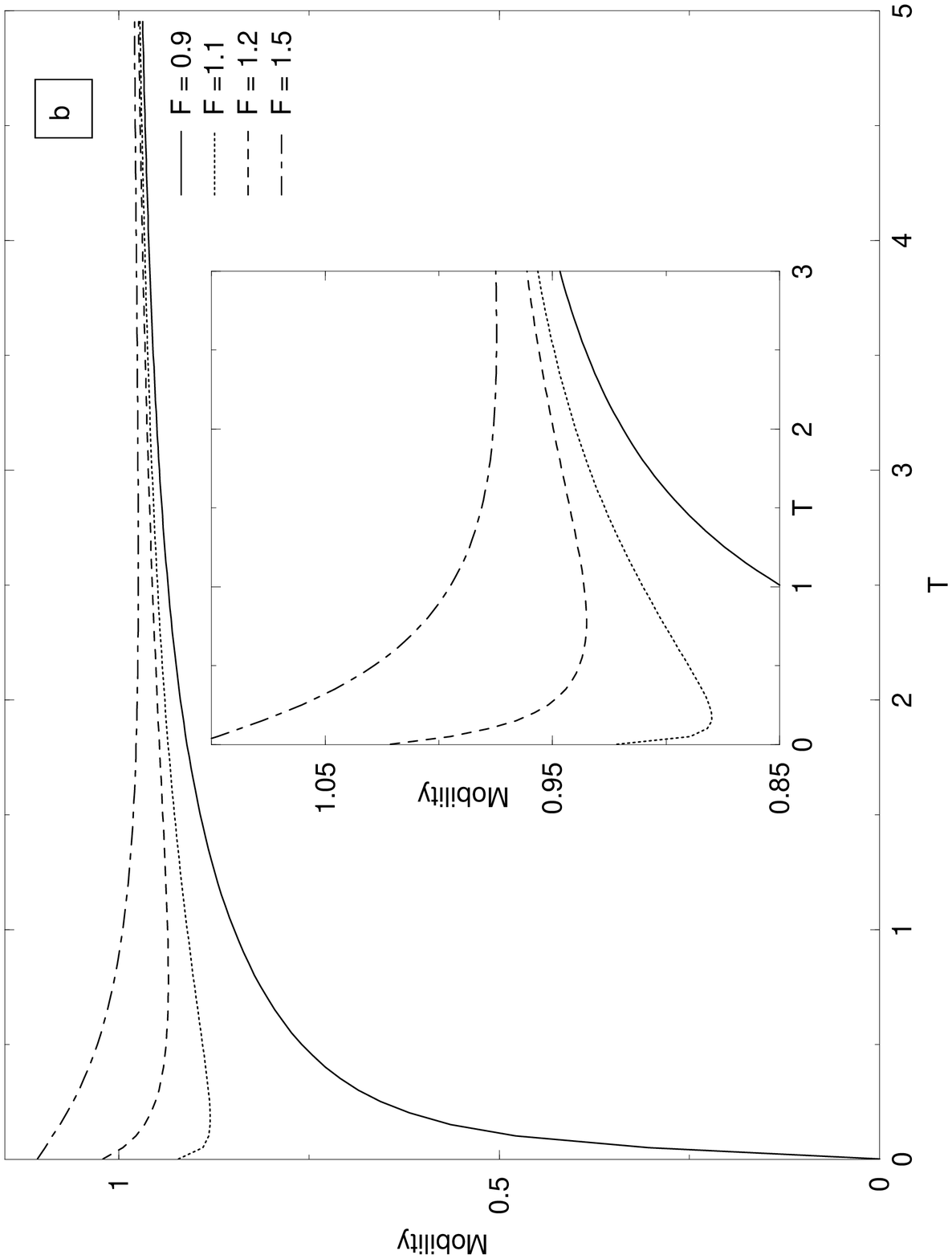}}
  \vspace{.5in}
  \caption{}
\end{figure}

\end{document}